\documentclass[12pt,a4paper]{article}

\usepackage[a4paper, top=2.5cm, bottom=2.5cm, left=2.5cm, right=2.5cm]{geometry}

\usepackage{amsmath,amssymb,amsthm}
\usepackage{graphicx}
\usepackage{subcaption}
\graphicspath{{./figures/}}

\theoremstyle{definition}
\newtheorem{theorem}{Theorem}
\newtheorem{definition}[theorem]{Definition}

\newtheorem{remark}[theorem]{Remark}

\usepackage{xifthen}
\newcommand{\textcite}[2][]{\ifthenelse{\isempty{#1}}{\cite{#2}}{\cite[#1]{#2}}}
\newcommand{\parencite}[2][]{\ifthenelse{\isempty{#1}}{\cite{#2}}{\cite[#1]{#2}}}

\begin{document}

\title{Non-Parametric Goodness-of-Fit Tests Using Tsallis Entropy Measures}
\author{Mehmet Sıddık Çadırcı$^1$}
\date{
	\slshape\small 
    $^1$ Faculty of Science, Department of Statistics, Cumhuriyet University, Sivas, Türkiye  \\[3ex] 
	\normalfont
}

\maketitle

\begin{abstract}
In this paper, we investigate new procedures for statistical testing based on Tsallis entropy, a parametric generalization of Shannon entropy. Focusing on multivariate generalized Gaussian and $q$-Gaussian distributions, we develop entropy-based goodness-of-fit tests based on maximum entropy formulations and nearest neighbour entropy estimators. Furthermore, we propose a novel iterative approach for estimating the shape parameters of the distributions, which is crucial for practical inference. This method extends entropy estimation techniques beyond traditional approaches, improving precision in heavy-tailed and non-Gaussian contexts. The numerical experiments are demonstrative of the statistical properties and convergence behaviour of the proposed tests. These findings are important for disciplines that require robust distributional tests, such as machine learning, signal processing, and information theory.

\end{abstract}

\section{Introduction}\label{sec:intro}
Entropy, being a fundamental concept in information theory, is a measure of the uncertainty that sits inside a kind of probability distribution. Entropy estimation finds applications in a number of fields, including statistical inference, cryptography, thermodynamics, and machine learning. The classical notion of entropy was given by Shannon~\cite{shannon1948mathematical} who defined the differential entropy of a continuous random vector with density $f: \mathbb{R}^m \rightarrow \mathbb{R}$ as
\begin{equation}\label{eq:diff_entropy}
H(f) = -\int_{\mathbb{R}^{m}} f(x) \log f(x)\, dx.
\end{equation}

To capture more flexible notions of uncertainty, several generalizations of Shannon entropy exist. One such generalization is R\'enyi entropy~\cite{nielsen2011r}, which introduces a tunable parameter to the entropy definition so one sort of distribution can be emphasized against another depending on whether the tail behavior is more or less important. For a random variable $X \in \mathbb{R}^m$ with density $f$, R\'enyi entropy is defined as
\begin{equation}\label{eq:renyi_entropy}
H^{*}_q(f) = \frac{1}{1 - q} \log \int_{\mathbb{R}^m} f^q(x)\, dx, \quad q \neq 1.
\end{equation}
As $q \to 1$, $H^*_q(f)$ asymptotically converges to the Shannon entropy in ~\eqref{eq:diff_entropy}.
Another important extension is Tsallis entropy~\cite{tsallis1988possible}, which attracted the attention of many researchers due to its implications in non-extensive thermodynamics, statistical mechanics, information geometry, and image analysis. The Tsallis entropy of a density $f$ is given by

\begin{equation}\label{eq:tsallis_entropy}
H_q(f) = \frac{1}{1 - q} \left( \int_{\mathbb{R}^{m}} f^q(x)\, dx - 1 \right), \quad q \neq 1.
\end{equation}
While Tsallis entropy is structurally similar to the R\'enyi entropy, it does not satisfy the conventional additivity property of the Shannon entropy. It follows a form of pseudo-additivity expressed as:
\begin{equation*}
H_q(X, Y) = H_q(X) + H_q(Y) + (1 - q) H_q(X) H_q(Y), 
\end{equation*}
for independent $X$ and $Y$.
A majority of recent studies have extended the theoretical and applied foundations of Tsallis entropy in various settings~\cite{dos1997generalization, alomani2023further, sati2015some, kumar2016some, kumar2017characterization, bulinski2019, bulinski2019a, berrett2019a, berrett2019b}. Consideration in statistical hypothesis testing frameworks is quite underdeveloped.

In this paper, goodness-of-fit tests have been proposed for multivariate generalized Gaussian and $q$-Gaussian distributions based on Tsallis entropy-type criteria. Combining the maximum entropy principle with $k$-nearest neighbor-based non-parametric estimation, we obtain flexible and data-driven test statistics. We study their consistency and asymptotic properties under various distributional assumptions.

The rest of the paper is organized as follows.
Section~\ref{sec:intro} deals with the conceptual background and motivation, defining Tsallis entropy and relating it with Shannon and Rényi entropy.
In Section~\ref{sec:maxent}, the maximum entropy principle and multivariate exponential power distributions are set out, thus providing the setting for our framework.
In Section~\ref{sec:tsallis_estimation}, we provide a class of nearest-neighbor-based nonparametric estimators for Tsallis entropy and study its consistency under general assumptions.
Section~\ref{sec:test_statistics} introduces entropy-based goodness-of-fit test statistics for $q$-Gaussian and generalized Gaussian distributions and discusses their asymptotic behaviour.
Section~\ref{sec:numerical_experiment} presents extensive Monte Carlo simulations of empirical distributions, convergence rates, and normal approximations for the test statistics proposed.
Finally, Section~\ref{sec:conclusion} discusses conclusions and prospects for future work, including estimations for dependent data and robust applications in machine learning.

\section{Principle of Maximum Entropy}\label{sec:maxent}
Let $X$ be a random vector in $\mathbb{R}^m$ with density $f(x)$ relative to the Lebesgue measure  on $\mathbb{R}^m$. We denote the set $S = \{x \in \mathbb{R}^m : f(x) > 0\}$ as the support of this distribution. The $q$-order Tsallis entropy, for some $q \in (0,1) \cup (1,\infty)$, is defined as follows:

\begin{equation}\label{eq:tsallis_entropy_2}
H_q(f) = \frac{1}{1 - q} \left(\int_S f^q(x) \, dx - 1\right), \quad q \neq 1.
\end{equation}

This entropy function is continuous and monotonically non-increasing concerning $q$. In the limit, the behaviour depends on the Lebesgue measure $|S|$ of the support. When $|S|$ is finite,
\[
\lim_{q \to 0} H_q(f) = \log |S|,
\]
but if $|S|$ is infinite, then the entropy diverges as $q$ goes to 0. Additionally,
\[
\lim_{q \to 1} H_q(f) = H(f) = -\int_S f(x) \log f(x) \, dx,
\]
which recovers the Shannon entropy.
Consider a location parameter $\alpha \in \mathbb{R}^m$ and a symmetric and positive definite covariance matrix $\Sigma \in \mathbb{R}^{m \times m}$. Then, the multivariate exponential power distribution $\text{MEP}_m(s, \alpha, \Sigma)$ is defined as follows:~\cite{solaro2004random}

\begin{equation}\label{eq:mep}
f(x; m, s, \alpha, \Sigma) = \frac{\Gamma(m/2+1)}{\pi^{m/2} \Gamma(m/s + 1) 2^{m/s} \sqrt{\det \Sigma}} \, \exp\left(-\frac{1}{2} \left[(x - \alpha)^T \Sigma^{-1} (x - \alpha)\right]^{s/2}\right),
\end{equation}

where $s > 0$ is a shape parameter governing the tail heaviness and peakedness of the distribution. The variance-covariance structure is given by $\text{Var}(X) = \beta \Sigma$, with scale factor
\begin{equation}\label{eq:mep-scalefactor}
\beta(m, s) = \frac{2^{2/s} \Gamma\left[(m+2)/s\right]}{m \Gamma(m/s)}.
\end{equation}

For $s = 2$, this distribution coincides with the multivariate normal distribution $\mathcal{N}(\alpha, \Sigma)$, while $s = 1$ yields the multivariate Laplace distribution. This family was originally introduced by~\cite{de1968estensione} and further examined by~\cite{kano1994consistency} and~\cite{gomez1998multivariate}. The $\text{MEP}_m$ class belongs to the elliptical family and includes symmetric Kotz-type distributions~\cite{fang1990}.

A special case arises when $\alpha = \mathbf{0}$ and $\Sigma = I_m$, the identity matrix. This yields the isotropic exponential power distribution $\text{IEP}_m(s)$ with density
\begin{equation}\label{eq:sep}
f(x; m, s) = \frac{\Gamma(m/2 + 1)}{\Gamma(m/s + 1) \pi^{m/2} 2^{m/s}} \, \exp\left(-\frac{1}{2} \|x\|^s\right),
\end{equation}
where $x \in \mathbb{R}^m$ and $\|x\|$ denotes the standard Euclidean norm. This simplification provides analytical tractability and is often used for simulation and theoretical derivations.

\section{Tsallis Entropy}\label{sec:tsallis_estimation}
Tsallis entropy generalizes the concept of Shannon entropy. It tends to apply to non-extensive systems, such as systems showing long-range interactions or non-Markovian dynamics. Its expression, acting on the probability density function $f : \mathbb{R}^m \to \mathbb{R}$, is as follows:

\[
    S_q(f) = \frac{1}{q - 1} \left(1 - \int_{\mathbb{R}^m} f^q(x) \, dx \right), \quad q \neq 1.
\]

This definition allows to recover the  Shannon entropy in the limit  $q \to 1$.

\subsection{Generalized Gaussian Distributions under Tsallis Entropy}
 Let $f^{\mathrm{GG}}(x)$ represent the probability density function for the multivariate Generalized Gaussian distribution  defined in \cite{cadirci2022entropy} as :
\[
    f^{\mathrm{GG}}(x) = \frac{1}{C(m,s,\Sigma)} \exp\left(-\frac{1}{2} h(x,\mu,\Sigma)^s\right),
\]
where $h(x,\mu,\Sigma) = (x - \mu)^T \Sigma^{-1} (x - \mu)$ is the Mahalanobis distance raised to the power $s$, and $C(m,s,\Sigma)$ is the normalization constant:

\[
    C(m,s,\Sigma) = \int_{\mathbb{R}^m} \exp\left(-\frac{1}{2} h(x,\mu,\Sigma)^s\right) \, dx.
\]

Then the integral for the Tsallis entropy becomes:

\[
    \int_{\mathbb{R}^m} [f^{\mathrm{GG}}(x)]^q \, dx = \frac{C(m,sq,\Sigma)}{C(m,s,\Sigma)^q},
\]

and the Tsallis entropy of multivariate Generalized Gaussian distribution is:

\[
    H_q(f^{\mathrm{GG}}) = \frac{1}{1 - q} \left(\frac{C(m,sq,\Sigma)}{C(m,s,\Sigma)^q} - 1\right).
\]

This expression is obtained and optimized in \cite{furuichi2009maximum}.

\subsection{The {q}-Exponential and {q}-Gaussian Distributions}
Consider the q-exponential function, a nonlinear generalization of the classical exponential function. 

\[
    \exp_q(x) = \begin{cases}
        [1 + (1 - q)x]^{1 / (1 - q)}, & \text{if } 1 + (1 - q)x > 0, \\
        0, & \text{otherwise}.
    \end{cases}
\]

The classical exponential function is recovered in the limit $q \to 1$. Another name for the one-dimensional $q$-Gaussian distribution is the $q$-exponential distribution.

\[
    f(x; a, \sigma, q) = C_q \left[1 - (1 - q) \frac{(x - a)^2}{2 \sigma^2}\right]_+^{1 / (1 - q)},
\]

where  $C_q$ is the normalization constant, $a$ is the position, and $\sigma > 0$ is the scale parameter. Depending on $q$, it extends the Gaussian family to the heavy-tailed or compactly supported cases.

\subsection{Multivariate {$q$}-Gaussian and Its Entropy}
The multivariate $q$-Gaussian distribution extends the above $\mathbb{R}^m$ distribution as follows:

\[
    f(x; \mu, \Sigma, q) = C_q \left[1 - (1 - q) \frac{(x - \mu)^T \Sigma^{-1} (x - \mu)}{2} \right]_+^{1 / (1 - q)}.
\]

Here $\mu \in \mathbb{R}^m$ is the mean, $\Sigma \in \mathbb{R}^{m \times m}$ corresponds to the covariance matrix, and $C_q$ provides the appropriate normalization. Define the change of variables as $y = \Sigma^{-1/2}(x - \mu)$, so that

\[
    f(y) \propto \left[1 - (1 - q) \frac{\|y\|^2}{2}\right]_+^{1 / (1 - q)}.
\]

The Tsallis entropy of the multivariate $q$-Gaussian distribution $G_{m,q}(\mu\ \Sigma)$ with mean vector $\mu \in \mathbb{R}^m$, covariance matrix $\Sigma \in \mathbb{R}^{m \times m}$, and  and entropic parameter $q$,  then becomes:

\[
    H_q(G_{m,q}(\mu\ \Sigma)) = \frac{1}{q - 1} \left(1 - C_q^q |\Sigma|^{\frac{1 - q}{2}} \int_{\mathbb{R}^m} \left[1 - (1 - q) \frac{\|y\|^2}{2}\right]_+^{q / (1 - q)} \, dy\right).
\]

Changing to spherical coordinates, the radial component is evaluated as follows:

\[
    \int_0^\infty \left[1 - (1 - q) \frac{r^2}{2}\right]_+^{q / (1 - q)} r^{m - 1} \, dr,
\]

which gives a closed form about the Beta and Gamma functions:

\[
    H_q(G_{m,q}(\mu\ \Sigma)) = \frac{1}{q - 1} \left(1 - C_q^q |\Sigma|^{\frac{1 - q}{2}} \frac{2^{m/2} \Gamma(m/2)}{(1 - q)^{m/2} \Gamma\left(\frac{m}{2} + \frac{1}{q - 1}\right)}\right).
\]

The dependence of entropy on $q$, $m$, and the geometry encoded in $\Sigma$ is highlighted by this expression.

\section{Tsallis Entropy: Statistical Estimation Method}

In this section,  we focus on the non-parametric estimation of Tsallis entropy for continuous distributions. Following \cite{martinez2000tsallis, abe2001heat, suyari2006unique}, we consider a $k$-NN estimator that avoids density estimation of Tsallis entropy. Let $X$ be a random vector in $\mathbb{R}^m$ with a Lebesgue-continuous density $f$. Given $N$ independent realizations $\mathcal{X}_N = \{X_1, \ldots, X_N\}$ drawn from $f$, the goal is to estimate:

\[
S_q(f) = \frac{1}{q - 1} \left( 1 - \int_{\mathbb{R}^m} f^q(x) \, dx \right), \quad q \neq 1.
\]

For $k \geq 1$ and $N > k$, let $\rho_{i,k,N}$ denote the Euclidean distance between $X_i$ and its $k$-th nearest neighbor in $\mathcal{X}_N \setminus \{X_i\}$. Then the estimator introduced in \cite{leonenko2008} is given by:
\begin{equation}\label{eq:estS_q}
\hat{S}_{k,N,q} = \frac{1}{N} \sum_{i=1}^{N} \left(\zeta_{i,k,N}\right)^{1 - q},
\end{equation}
where
\[\zeta_{i,k,N} = (N - 1) C_k V_m \rho_{i,k,N}^m, \quad C_k = \left[ \frac{\Gamma(k)}{\Gamma(k + 1 - q)} \right]^{1 / (1 - q)},\]
and $V_m = \pi^{m/2} / \Gamma(m/2 + 1)$ is the volume of the $m$-dimensional unit ball.

\begin{definition}\label{def:r-moment}
For any positive integer  $r$, the \emph{$r$-moment} of $f$ under Tsallis entropy is:
\[K_r(f) = \mathbb{E}(\|X\|^r) = \frac{1}{q - 1} \int_{\mathbb{R}^m} \|x\|^r f^q(x) \, dx.\]
The \emph{critical moment} is defined as follows:
\[r_c(f) = \sup \{ r > 0 : K_r(f) < \infty \}.
\]
This defines the maximal order of finite moments admissible under $f^q$.
\end{definition}

\begin{remark}
A Monte Carlo approximation to $S_q(f)$, assuming knowledge of $f$, is:
\[\frac{1}{N} \sum_{i=1}^{N} f^{q - 1}(X_i).\]
\end{remark}

The estimator $\hat{S}_{k,N,q}$ may be interpreted as a plug-in estimator based on a $k$-NN density estimator:
\[\hat{S}_{k,N,q} = \frac{1}{N} \sum_{i=1}^{N} \left[ \hat{f}_{N,k}(X_i) \right]^{q - 1}, \quad \hat{f}_{k,N}(x) = \frac{1}{(N - 1) C_k V_m \rho_{k+1,N}(x)^m}.
\]
This closely resembles the non-parametric estimator proposed in \cite{loftsgaarden1965nonparametric} and generalized in \cite{devroye1977strong}.

We assume $X_1, \ldots, X_N$ are i.i.d. samples from a distribution $\mu$ with a density $f$ plus possibly a finite number of singular components. In such settings, zero-distance degeneracy can be avoided, and the estimator \eqref{eq:estS_q} retains asymptotic consistency for the continuous component $f$.

\begin{theorem}[\cite{siddik2021statistical, cadirci2021entropy}]\label{thm:main}
Fix $q \in (0,1)$ and $k \geq 1$. Then:
\begin{enumerate}
\item If $S_q(f) < \infty$ and
\begin{equation}\label{eq:condition-conv-in-mean}
r_c(f) > \frac{m(1 - q)}{q},
\end{equation}
then \( \mathbb{E}(\hat{S}_{k,N,q}) \to S_q(f) \) as $N \to \infty$.

\item If additionally $q > 1/2$ and
\begin{equation}\label{eq:condition-conv-in-mean-square}
r_c(f) > \frac{2m(1 - q)}{2q - 1},
\end{equation}
then \( \mathbb{E}\left[\left(\hat{S}_{k,N,q} - S_q(f)\right)^2\right] \to 0 \) as $N \to \infty$.
\end{enumerate}
\end{theorem}

\begin{remark}
For $q \in (1, (k+1)/2)$, it was shown in \cite{leonenko2008} that the same consistency results hold under appropriate moment conditions.
\end{remark}

\begin{remark}
If $f(x) = O(|x|^{-\beta})$ as $|x| \to \infty$ for some $\beta > m$ and $q \in (0,1)$, then $r_c(f) = \beta - m$, ensuring the condition \eqref{eq:condition-conv-in-mean} is satisfied. For related discussions, see \cite{penrose2011}.
\end{remark}

\section{Test Statistics and Hypothesis Testing for \( T(x, a, \sigma) \)}\label{sec:test_statistics}
Let \( \mathcal{K} \) denote a class of distributions for which the $k$-nearest neighbor entropy estimator $\hat{S}_{k,N,q}$ satisfies, for any fixed $k \geq 1$ and $q > 0.5$:
\begin{align*}
\mathbb{E}(\hat{S}_{k,N,q}) &\longrightarrow S_q \quad \text{as } N \to \infty, \\[0.5ex]
\hat{S}_{k,N,q} &\longrightarrow S_q \quad \text{in probability as } N \to \infty.
\end{align*}
By Theorem~\ref{thm:main}, the distributions \( T^1(x; a, q, \sigma) \) and \( T^2(x; a, q, \sigma) \) are included in this class.

Consider now i.i.d. random vectors \( X_1, X_2, \ldots, X_N \sim f \in \mathcal{K} \). The sample covariance matrix is given by:
\[ \hat{S}_N = \frac{1}{N - 1} \sum_{i=1}^N (X_i - \bar{X})(X_i - \bar{X})^T, \quad \text{with } \bar{X} = \frac{1}{N} \sum_{i=1}^N X_i. \]

\subsection{Test Statistics}
The null hypothesis that $X$ follows either $T^1(x; a, q, \sigma)$ or $T^2(x; a, q, \sigma)$ is assessed by the following test statistics, which are defined as follows:

\begin{enumerate}
    \item For \( H_0: X \sim T^1(x; a, q, \sigma) \), with $q \in (1, 3)$, define:
    \begin{equation}\label{eq:tstat-student}
    Q^{\mathrm{Tsallis}}_{N,k}(m, q) = H_q^{\mathrm{upper}} - \hat{S}_{k,N,q},
    \end{equation}
    where $H_q^{\mathrm{upper}} = \frac{1}{2} \log |\hat{\Sigma}_N| + T^1(x; a, q, \sigma)$ denotes the maximum Tsallis entropy under the assumed model.

    \item For \( H_0: X \sim T^2(x; a, q, \sigma) \), with $q \in (0, 1)$, define:
    \begin{equation}\label{eq:tstat-pearsonII}
    Q^{\mathrm{Tsallis}^*}_{N,k}(m, q) = H_q^{\mathrm{upper}} - \hat{S}_{k,N,q},
    \end{equation}
    where $H_q^{\mathrm{upper}} = \frac{1}{2} \log |\hat{\Sigma}_N| + T^2(x; a, q, \sigma)$.
\end{enumerate}

\subsection{Asymptotic Behavior}
According to the law of large numbers, we have that \( \hat{S}_N \xrightarrow{P} \Sigma \). Furthermore, according to Theorem~\ref{thm:main}, we have that \( \hat{S}_{k,N,q} \xrightarrow{P} S_q \). Using Slutsky's theorem, the test statistics converge in probability as:

\[ \lim_{N \to \infty} Q^{\mathrm{Tsallis}}_{N,k}(m, q) \xrightarrow{P}
\begin{cases}
0 & \text{if } X \sim T^1(x; a, q, \sigma), \\
c > 0 & \text{otherwise},
\end{cases} \]
\[ \lim_{N \to \infty} Q^{\mathrm{Tsallis}^*}_{N,k}(m, q) \xrightarrow{P}
\begin{cases}
0 & \text{if } X \sim T^2(x; a, q, \sigma), \\
c > 0 & \text{otherwise},
\end{cases} \]
where $c$ is a constant. It depends on the divergence between $f$ and the assumed distribution.

\section{Numerical experiments} \label{sec:numerical_experiment}

The extensive numerical evaluations presented here are motivated by the complexity inherent in deriving analytical null distributions for the proposed Tsallis entropy-based test statistics. Specifically, analytical tractability is hindered by intricate dependencies. These dependencies are in the $k$-nearest neighbor ($k$-NN) estimators. Monte Carlo simulations are used to empirically evaluate test statistics, providing insights into their performance under various configurations.

\subsection*{Challenges in Null Distribution}

The exact analytic null distributions of the test statistics \( Q^{\mathrm{Tsallis}}_{N,k} \) and \( Q^{\mathrm{Tsallis}^*}_{N,k} \) 
cannot be generated because of  the complex dependence between the distance measures and density estimates used for 
$k$-NN entropy estimation. Although asymptotic and central limit theorems have been introduced in the previous works \cite{penrose2011, delattre2017, berrett2019} , these analytical schemes do not sufficiently consider  the highly complex dependence structures inherent to entropy estimators. Therefore, Monte Carlo simulations naturally provide a solution to the performance evaluation of the proposed statistics.

\subsection*{Stochastic Generation of \( q \)-Gaussian Samples}

An accurate numerical evaluation of the Tsallis entropy-based tests requires robust methods for sampling from multivariate q-Gaussian distributions, denoted \( q\text{-}\mathcal{G}(m, q, \Sigma) \). Direct sampling is challenging because of the introduced nonlinearity of the parameter \( q \). However, a stochastic approach allows for efficient and precise generation of samples.

Specifically, we first generate a standard Gaussian vector, \(\mathbf{Z} \sim \mathcal{N}_m(\mathbf{0}, \mathbf{I})\). Then, we independently define a scalar random variable \( R \) with the following density:
\[
f_R(r) \propto \left[1 - (1 - q)\frac{r^2}{2}\right]_+^{\frac{1}{1 - q}}.
\]

Combining these elements yields the multivariate \( q \)-Gaussian random vector \( \mathbf{X} \in \mathbb{R}^m \) via:
\[
\mathbf{X} = \boldsymbol{\mu} + R \mathbf{\Sigma}^{1/2} \mathbf{Z},
\]

where \( \boldsymbol{\mu} \) is the mean vector and \( \mathbf{\Sigma}^{1/2} \) represents the Cholesky decomposition (matrix square root) of the covariance matrix \( \Sigma \).

The distribution of \( R^2 \) explicitly depends on the parameter \( q \): it follows a Beta distribution provided \( q < 1 \), while, for \( q > 1 \), it follows a Gamma distribution. Moreover, this distribution converges smoothly to the normal type when \( q \rightarrow 1 \), highlighting its suitability for comparative entropy-based analysis.

\begin{figure}[htbp]
    \centering
    \includegraphics[width=\textwidth]{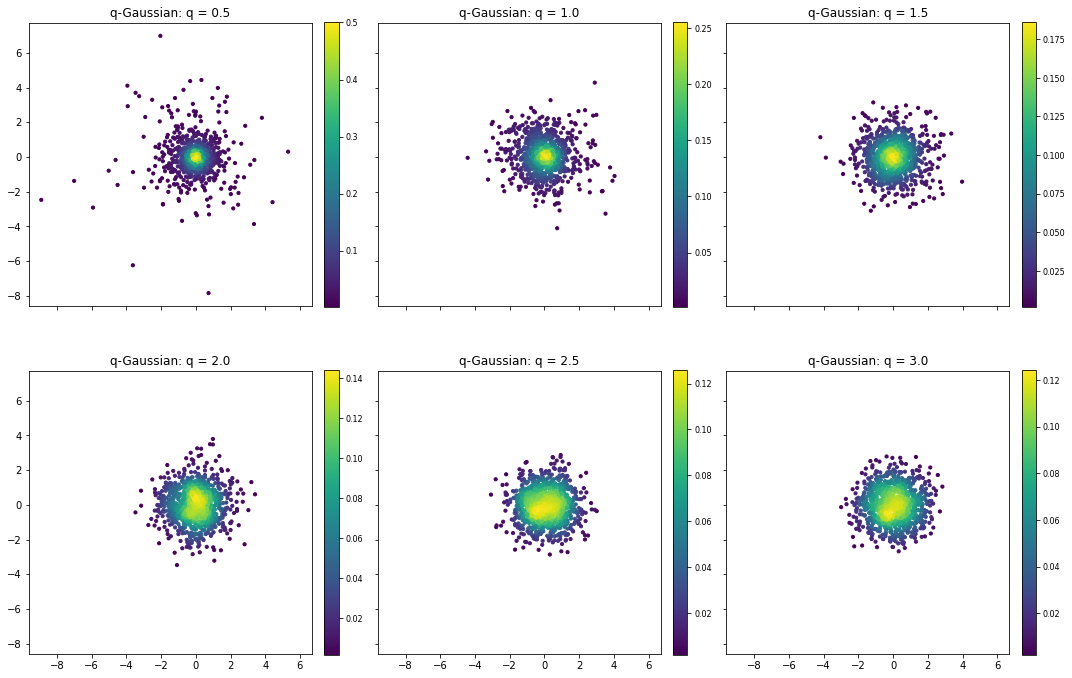}
    \caption{Scatter plots depicting simulated multivariate \( q \)-Gaussian samples in \( \mathbb{R}^2 \). Each subplot represents a different value of \( q \), illustrating clearly the differences in concentration patterns and tail behavior.}
    \label{fig:colored_point_MqG}
\end{figure}

To demonstrate these distributional aspects graphically, Figure~\ref{fig:colored_point_MqG} provides scatter plots in subplots falling under different values of \( q \), hence demonstrating the central concentration and tail behaviour's variability concerning \( q \).

\subsection*{Empirical Density and Analysis of Log-Density}

\begin{figure}[htbp]
\centering
\includegraphics[width=0.45\textwidth]{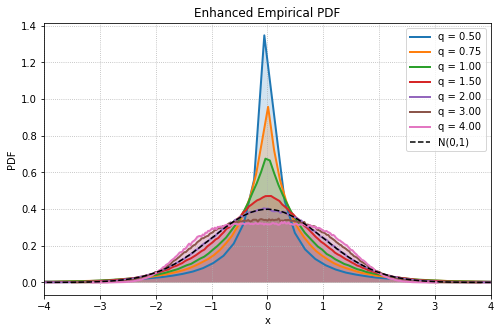}\hfill
\includegraphics[width=0.45\textwidth]{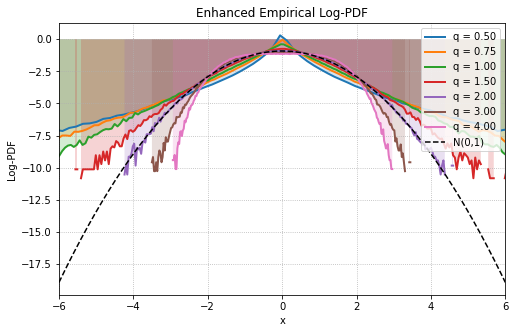}
\caption{Empirical pdf on the left and log-pdf for multivariate $q$-Gaussian samples of dimension $m = 1$ on the right. The figure on the left illustrates that decreasing $q$ sharpens the peak and widens the tails, while the figure on the right clearly depicts the deviation from Gaussian tails in log density space and heavier tails for lower values of $q$.}
\label{fig:EPDF_LOG_EPDF_MST}
\end{figure}

To elucidate the shape and tail characteristics of the multivariate $q$-Gaussian distribution, we simulate $N = 10^6$ samples from $q\text{-}\mathcal{G}(m, q, \Sigma)$ for various values of $q$. The resulting empirical probability density functions (PDFs) and their corresponding log-density plots are presented in Figure~\ref{fig:EPDF_LOG_EPDF_MST}.

The empirical PDFs establish quite conclusively that as $q$ decreases, the distributions behave more similarly to peak around the mean than a standard Gaussian distribution would, and the tails become enormously heavier. On the other hand, linear log-density plots enable a closer view of how higher-scale deviations deviate from Gaussianity, which is precisely what is associated with these heavier tails. These figures provide simple graphical evidence that reveals clearly how pronounced the $q$ effects are on the tail behaviour and the overall distribution shape.

\subsection*{Monte Carlo Study of Test Statistic Behaviour}

\begin{figure}[htbp]
\centering
\includegraphics[width=\textwidth]{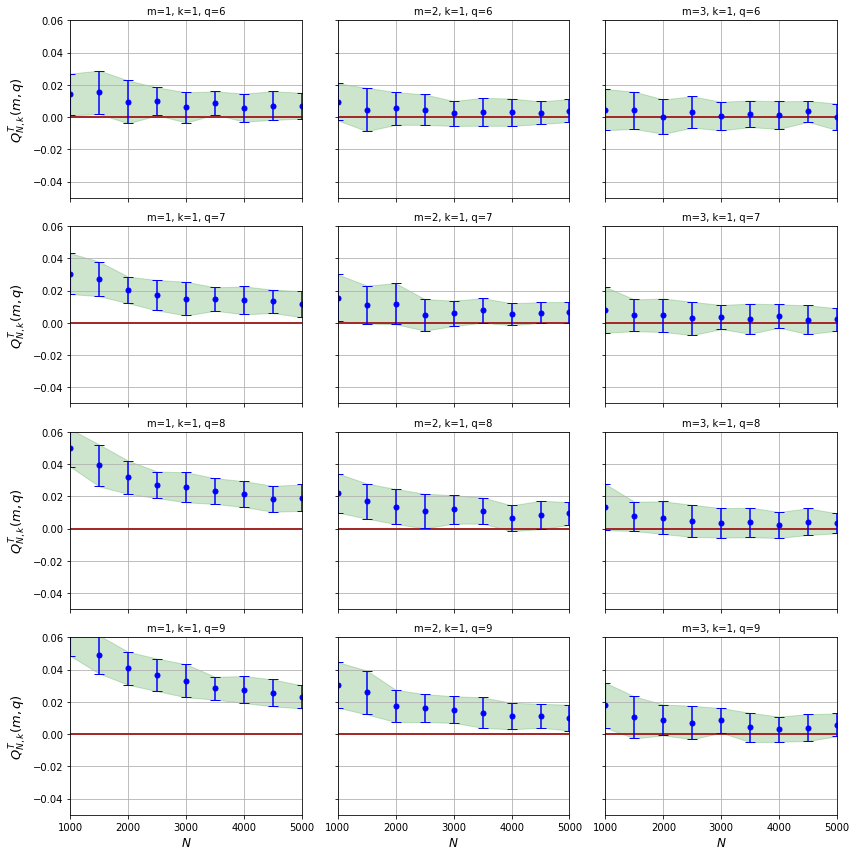}
\caption{The convergence behavior of $Q^T_{N,k}(m,q)$ for neighborhood size $k = 1$. The figure illustrates clearly the decreasing variance with increasing sample size and the convergence to theoretical expectations.}
\label{fig:cons-MqD_cons_k_1}
\end{figure}

\begin{figure}[htbp]
\centering
\includegraphics[width=\textwidth]{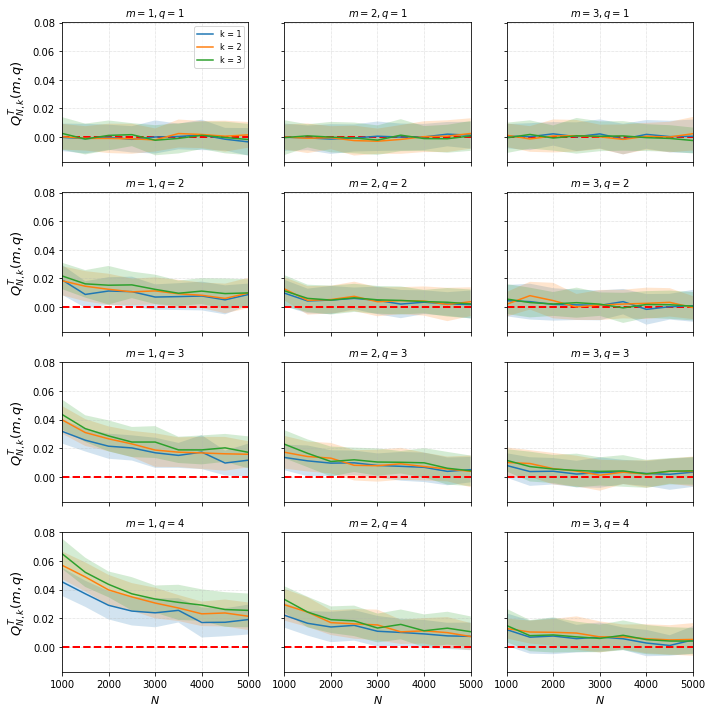}
\caption{Consistency of $Q^T_{N,k}(m,q)$ values across neighborhood sizes ($k = 1, 2, 3$). Error bars denote standard deviations, showing that while uncertainty declines with increasing sample size, the parameter remains fairly stable against $q$ and $m$.}
\label{fig:cons-MqD_cons_k_all}
\end{figure}

We used extensive Monte Carlo simulations to study the convergence behavior of the proposed test statistic, $Q^T_{N,k}(m,q)$. For each $q$  parameter set and $m$ parameter size, we performed $M = 100$ replications for various sample sizes between $N = 500$ and $N = 5000$. As shown in Figure~\ref{fig:cons-MqD_cons_k_1}, the test statistic converges for $k = 1$ and variability decreases as the sample size increases.Figure~\ref{fig:cons-MqD_cons_k_all} extends this analysis to neighborhood sizes$k = 1, 2$, and $3$ and shows that the test statistic is consistent and stable across multiple dimensions and parameters.

\subsection*{Violin Plots and Distributional Analysis}

\begin{figure}[htbp]
\centering
\includegraphics[width=\textwidth]{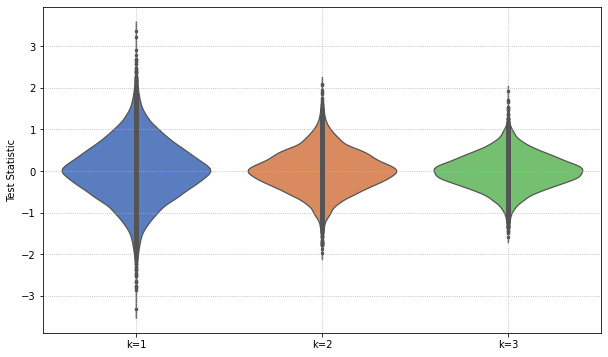}
\caption{The violin plots illustrate the empirical distributions of $Q^T_{N,k}(m,q)$ for a dimension of $m = 2$ and neighborhood sizes of $k = 1, 2$, and $3$. As $q$ approaches Gaussianity, distributional symmetry and narrowing variance are evident.}
\label{fig:violin_MqG}
\end{figure}

The violin plots in Figure ~\ref{fig:violin_MqG} depict the empirical distributions of $Q^T_{N,k}(m,q)$ for dimensions $m=2$. These plots clearly demonstrate distributional shifts toward symmetry and a reduction in variance when the parameter $q$ approaches unity. These visualizations intuitively represent the nuances of the distributions generated by the proposed test statistics.

\subsection*{Plot of Q-Q for Normality Check}

\begin{figure}[htbp]
\centering
\includegraphics[width=\textwidth]{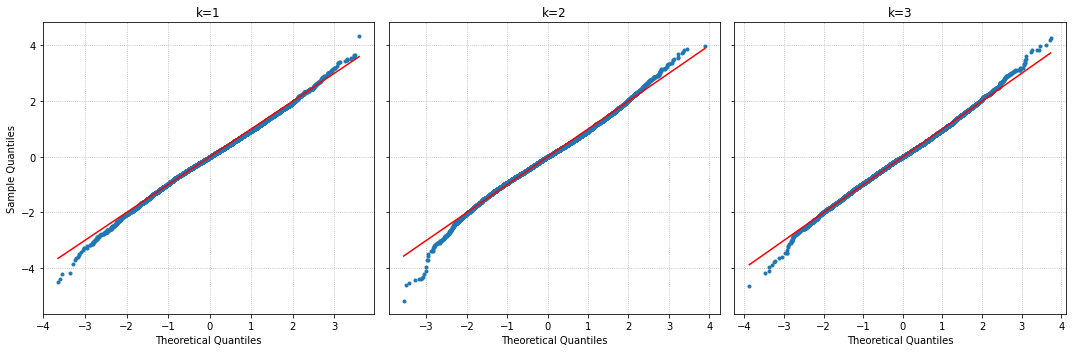}
\caption{Q-Q plots of $Q^T_{N,k}(m,q)$  are compared with standard Gaussian quantities for dimension $m = 2$. The KDE-enhanced visualization verifies convergence to normality with decreasing $q$.}
\label{fig:q_q_MqG}
\end{figure}

Kernel density estimation and Q-Q plots were also employed to further assess the accuracy of the normal approximation. Figure~\ref{fig:q_q_MqG} demonstrates the progressive alignment of the empirical distribution with Gaussian quantities as the parameter $q$ approaches unity. This graphical verification is a strong confirmation of the normality assumption under these conditions and provides further confidence in the theoretical robustness of our procedure.

\subsection{Empirical Distribution of the Test Statistics}
\label{subsec:empirical-dist}

We perform a detailed simulation-based analysis to investigate the limiting distribution of test statistics \( Q^T_{N,k}(m,q) \) under various \( (N, k, m, q) \) configurations. For each configuration, we generate \( n = 100 \) independent samples of size \( N \) from the distribution \( q\text{-}\mathcal{G}(m, q, \Sigma) \) and calculate the corresponding test statistic \( Q^T_{N,k}(m,q) \).

\begin{figure}[htbp]
\centering
\includegraphics[width=\textwidth]{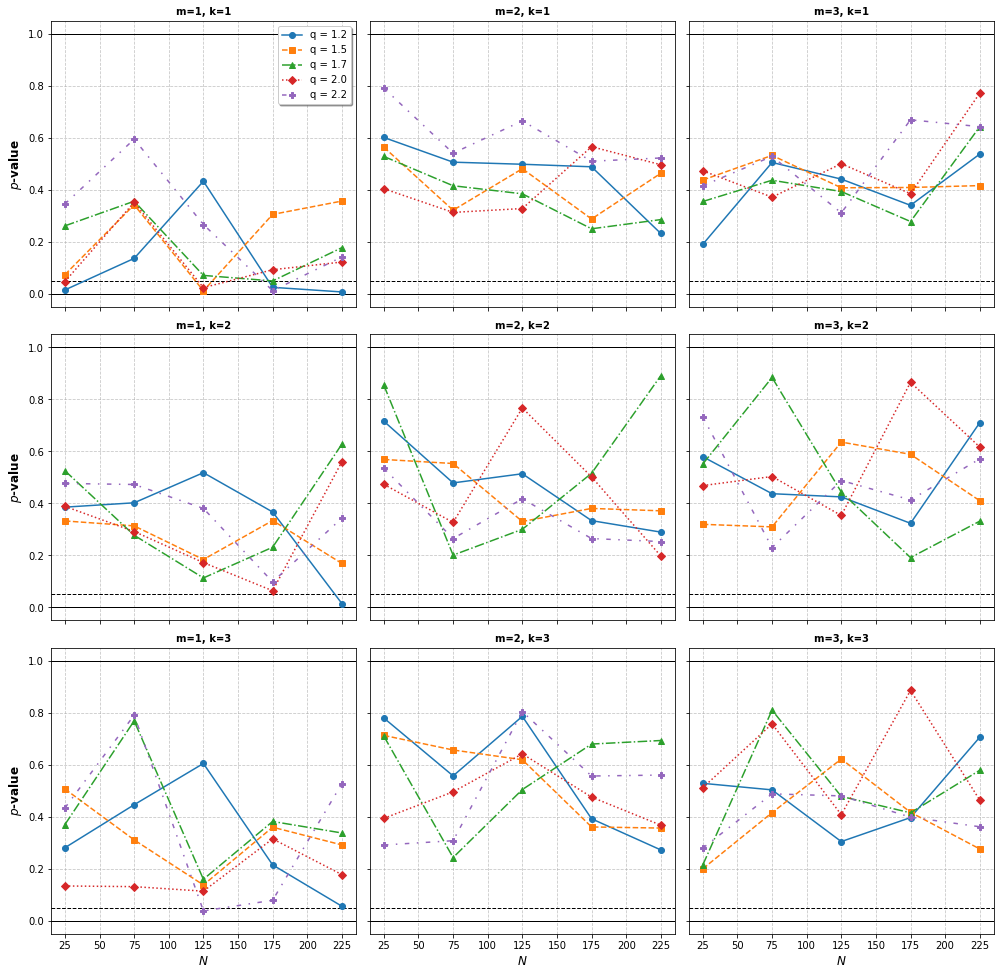}
\caption{Average Shapiro-Wilk \( p \)-values averaged for \( Q^T_{N,k}(m,q) \) versus sample size \( N \). The points are averaged over \( M = 1000 \) replicates, highlighting the enhanced normality as \( q \to 1 \). Larger neighborhood sizes (\( k \)) indicate a slight decrease in the normality.}
\label{fig:shapiro}
\end{figure}

The Shapiro-Wilk test~\cite{shapiro1965} is applied to ascertain normality for each set of 100 statistic values. The procedure is repeated \(M=1000\) times to protect against weak evidence from random variability. Figure~\ref{fig:shapiro} depicts the averaged Shapiro-Wilk \(p\)-value across repetitions, quite clearly showing that as \(q \to 1\), the distribution of \(Q^T_{N,k}(m,q)\) tends to normal. Interestingly, the normal approximation becomes slightly weaker as the neighborhood size \(k\) increases.

Furthermore numerical applications

 are proof of the theoretical idea that \[
\mathbb{E}[Q^T_{N,k}(m,q)] \to 0 \quad \text{as} \quad N \to \infty,
\]
and thus proves the consistency of the proposed test.

Empirically this is achieved by determining the \(\alpha\)-quantile \( \bar{q}_{\alpha} \) of \( Q^T_{N,k}(m,q) \) such that
\[
\mathbb{P}\left(Q^T_{N,k}(m,q) > \bar{q}_{\alpha}\right) = \alpha.
\]

Table~\ref{table:crt_val_MST} presents these critical results for the \( \alpha = 0.05 \) significance level, calculated from \( M = 1000 \) replicates.

\begin{table}[ht]
\centering
\renewcommand\arraystretch{1.2}
\caption{Critical values of the test statistics $Q^T_{N,k}(m,q)$ at $\bar{q}_{0.05}$ for the 5\% significance level. Estimated by using $M = 1000$ Monte Carlo replications.}
\begin{tabular}{ccccc|ccc}
\hline
$q$ & N & \multicolumn{3}{c}{m=2} & \multicolumn{3}{c}{m=3} \\ 
\cline{3-5} \cline{6-8}
& & k=1 & k=2 & k=3 & k=1 & k=2 & k=3 \\
\hline
1.2 & 100  & 0.03127 & 0.03000 & 0.02659 & 0.03072 & 0.02898 & 0.02969 \\
    & 200  & 0.03181 & 0.03051 & 0.02905 & 0.03056 & 0.03182 & 0.02725 \\
    & 300  & 0.02874 & 0.03187 & 0.03063 & 0.02902 & 0.02562 & 0.03294 \\
    & 400  & 0.03118 & 0.02641 & 0.03128 & 0.02883 & 0.02660 & 0.03045 \\
    & 500  & 0.03300 & 0.02898 & 0.02929 & 0.02796 & 0.03089 & 0.03099 \\
    & 600  & 0.03064 & 0.02871 & 0.03097 & 0.02968 & 0.02909 & 0.02714 \\
    & 700  & 0.02996 & 0.03009 & 0.02476 & 0.03131 & 0.03267 & 0.02884 \\
    & 800  & 0.03019 & 0.02684 & 0.03048 & 0.02993 & 0.02962 & 0.02895 \\
    & 900  & 0.03177 & 0.02882 & 0.03261 & 0.03034 & 0.03099 & 0.02893 \\
    & 1000 & 0.03079 & 0.03206 & 0.02881 & 0.03010 & 0.02946 & 0.02978 \\
1.5 & 100  & 0.02912 & 0.03386 & 0.02988 & 0.02988 & 0.02828 & 0.02485 \\
    & 200  & 0.02905 & 0.02997 & 0.03007 & 0.02933 & 0.02865 & 0.03134 \\
    & 300  & 0.02852 & 0.02648 & 0.03056 & 0.03181 & 0.02937 & 0.02675 \\
    & 400  & 0.02734 & 0.03029 & 0.02917 & 0.03009 & 0.02674 & 0.02943 \\
    & 500  & 0.03059 & 0.02816 & 0.03036 & 0.03042 & 0.03302 & 0.03244 \\
    & 600  & 0.03115 & 0.02998 & 0.03105 & 0.03163 & 0.03039 & 0.02986 \\
    & 700  & 0.03164 & 0.03099 & 0.02826 & 0.03181 & 0.02947 & 0.02999 \\
    & 800  & 0.03029 & 0.02781 & 0.03065 & 0.02941 & 0.02927 & 0.02920 \\
    & 900  & 0.02849 & 0.03079 & 0.02913 & 0.02989 & 0.03007 & 0.02513 \\
    & 1000 & 0.03143 & 0.02950 & 0.03023 & 0.03059 & 0.03115 & 0.02866 \\
2.5 & 100  & 0.02811 & 0.02875 & 0.02882 & 0.02872 & 0.03005 & 0.03042 \\
    & 200  & 0.02976 & 0.02930 & 0.02910 & 0.03115 & 0.02961 & 0.03185 \\
    & 300  & 0.03042 & 0.02860 & 0.02629 & 0.03276 & 0.02899 & 0.03013 \\
    & 400  & 0.03084 & 0.02722 & 0.02662 & 0.03138 & 0.03100 & 0.03037 \\
    & 500  & 0.03060 & 0.02875 & 0.02885 & 0.02939 & 0.03217 & 0.02747 \\
    & 600  & 0.02972 & 0.02692 & 0.03030 & 0.03267 & 0.03116 & 0.02934 \\
    & 700  & 0.02947 & 0.03107 & 0.02988 & 0.02935 & 0.03057 & 0.03037 \\
    & 800  & 0.02583 & 0.03158 & 0.02981 & 0.03081 & 0.03048 & 0.02846 \\
    & 900  & 0.02913 & 0.03029 & 0.02897 & 0.03186 & 0.02851 & 0.03245 \\
    & 1000 & 0.02899 & 0.03079 & 0.03024 & 0.02940 & 0.02842 & 0.03047 \\
\hline
\end{tabular}
\label{table:crt_val_MST}
\end{table}

We also estimate the speed of convergence by applying the following regression model:
\[
\log |\mathbb{E}\,\bar{Q}^T_{N,k}(m,q)| = \alpha_{m,q,k} + \beta_{m,q,k}\log N - \frac{1}{2}\log N.
\]
The values for the slope \( \beta_{m,q,k} \) are tabulated in Table~\ref{table:pest-qGaussian} to illustrate the dependence of the convergence rates on the parameters \( m \), \( k \) and \( q \). The smaller or more negative these slope values are, the slower the rate of convergence and stabilizes with values of (\( q \to 1 \)) as we approach Gaussianity.

\begin{table}[htbp]
\centering
\renewcommand\arraystretch{1.2}
\caption{Slope values $\beta$ in the log-log regression \( \log |\mathbb{E}\, \bar{Q}^T_{N,k}(m,q)| = \alpha_{m,q,k} + \beta_{m,q,k} \log N - \frac{1}{2}\log N \) for the multivariate $q$-Gaussian distribution.}
\begin{tabular}{c|ccc|ccc|ccc}
\hline
$q$  & \multicolumn{3}{c|}{$m = 1$} & \multicolumn{3}{c|}{$m = 2$} & \multicolumn{3}{c}{$m = 3$} \\
\cline{2-4} \cline{5-7} \cline{8-10}
& $k = 1$ & $k = 2$ & $k = 3$ & $k = 1$ & $k = 2$ & $k = 3$ & $k = 1$ & $k = 2$ & $k = 3$ \\
\hline
1.2   &  0.0085 & 0.0111 & 0.0093 & 0.0006 & 0.0004 & 0.0003 & 0.0004 & 0.0003 & 0.0003 \\
1.5   &  0.0047 & 0.0050 & 0.0045 & 0.0000 & 0.0001 & 0.0001 & 0.0000 & 0.0000 & 0.0000 \\
1.7   &  0.0015 & 0.0011 & 0.0014 &-0.0001 & 0.0001 & 0.0001 & 0.0001 & 0.0001 & 0.0001 \\
2.0   &  0.0005 & 0.0006 & 0.0006 & 0.0000 & 0.0000 & 0.0000 & 0.0000 & 0.0000 & 0.0000 \\
2.2   &  0.0007 & 0.0004 & 0.0004 & 0.0002 & 0.0002 & 0.0001 & 0.0000 & 0.0000 &-0.0001 \\
2.5   &  0.0002 & 0.0002 & 0.0002 &-0.0002 & 0.0000 & 0.0001 &-0.0001 &-0.0001 & 0.0000 \\
3.0   & -0.0004 &-0.0004 &-0.0001 &-0.0001 & 0.0000 & 0.0001 &-0.0001 & 0.0000 &-0.0001 \\
3.5   &  0.0002 & 0.0001 & 0.0002 & 0.0000 &-0.0001 &-0.0001 & 0.0000 & 0.0001 & 0.0001 \\
4.0   &  0.0003 & 0.0001 &-0.0001 & 0.0001 & 0.0003 & 0.0003 &-0.0001 &-0.0001 & 0.0000 \\
\hline
\end{tabular}
\label{table:pest-qGaussian}
\end{table}

Our simulations demonstrate that the proposed Tsallis entropy-based testing exhibits strong convergence properties with good tail precision compared to some classical entropy measures. In this paper, we emphasize the difficulty in choosing the neighbourhood parameter (\( k \)), which is perhaps the drawback of the method since the sensitivity of the test and the computational cost depend heavily on the choice of this parameter. An interesting line for further research could be to address this aspect by designing adaptive schemes for the choice of (\( k \)), thus increasing its practicality against large-scale data,
\clearpage
\section{Conclusion}
\label{sec:conclusion}

This paper has established a new class of statistical methods for testing goodness-of-fit with Tsallis entropy, particularly for multivariate generalized Gaussian and $q$-Gaussian distributions. We present several entropy-based test statistics based on nearest $k$-neighbor estimates and the maximum entropy principle. Such test statistics offer an alternative to traditional likelihood-based methods, especially when the usual assumptions, such as normality or light-tailedness, may be violated.

The paper contributes theoretically by developing and analyzing the convergence rates of a nonparametric estimator for Tsallis entropy with very strict moment-based conditions. The results are then applied to compactly supported ($q>1$) and heavy-tailed ($q<1$) distributions and utilized to derive test statistics. Asymptotic properties are formally proved, while issues related to the derivation of the full null distribution are addressed using high-resolution Monte Carlo simulations.

Extensive simulation studies have shown that the proposed statistics empirically converge to Gaussianity and are sensitive to deviations in $q$ and distribution shapes. The empirical quantities and critical values estimated for various parameter values provide a user-friendly toolkit for the practical application of entropy-based tests.

Future directions for research could include extending this procedure to hypothesis testing under dependence settings such as a time series or spatial model, and deriving bootstrap-based approximations for the null distribution. Another promising direction to explore is the potential connection between Tsallis entropy and robust machine learning models in the presence of heavy-tailed distributions.

In conclusion, Tsallis entropy offers a powerful perspective for statistical inference in non-exhaustive settings. The proposed framework contributes to the growing literature on entropy-based approaches with strong theoretical support but of great practical importance.



%
\clearpage
\bibliographystyle{plain}

\bibliography{tsallis} 


\end{document}